\documentclass[A4paper]{jpconf}
\usepackage{graphicx}
\newcommand\ro {\hat\rho}
\newcommand\Ho {\hat H}
\newcommand\Dc {\mathcal{D}}
\newcommand\la {\lambda}
\newcommand\dphi{\delta\phi}
\newcommand\psik{\vert\psi\rangle}
\newcommand\dd {\mathrm{d}}
\newcommand\rb {\mathbf{r}}
\newcommand\sx {\mathbf{s}}
\newcommand\pbo{\hat\mathbf{p}}
\newcommand\xbo{\hat\mathbf{x}}
\newcommand\xb {\mathbf{x}}
\newcommand\xba{\langle\mathbf{\xbo}\rangle}
\newcommand\pba{\langle\mathbf{\pbo}\rangle}
\newcommand\vb {\mathbf{v}}
\newcommand\fo {\hat f}
\newcommand\foa{\langle{\hat f}\rangle}
\newcommand\Fb {\mathbf{F}}

\begin{document}

\title{Does wave function collapse cause gravity? }

\author{Lajos Di\'osi}

\address{Research Institute for Particle and Nuclear Physics, H-1525 Budapest 114, P.O.Box 49, Hungary}

\ead{diosi@rmki.kfki.hu}

\begin{abstract}
We give a twist to the assumption - discussed in various earlier works - that
gravity plays a role in the collapse of the wave function. This time we
discuss the contrary assumption that the collapse of the wave function  
plays a role in the emergence of the gravitational field. We start from the 
mathematical framework of a particular Newtonian gravitational collapse theory 
proposed by the author longtime ago, and we reconciliate it with the classical
equivalence principle.
\end{abstract}

\section{Introduction}
For a suitable theory of quantum-gravity, one has to revise the standard 
concept of gravity or, alternatively, the standard concept of the quantum, or both.
Let us, e.g., consider the Wheeler-DeWitt \cite{Whe68,DeW67,Per68} equation of quantum-gravity:
\begin{equation}\label{WDW}
H\left(g,\frac{\partial}{\partial g},q,\frac{\partial}{\partial q}\right)\Psi(g,q)~=~0~,
\end{equation}
where $H$ and $\Psi$ are the quantized Hamilton function and the wave function of the universe,
resp., in function of the $3$-geometry $g$ and the matter fields $q$.
The generic solutions $\Psi(g,q)$  do not adapt any 4-geometry, only
the semi-classical approximate solutions do. The wave function $\Psi(g,q)$ does not possess
the standard statistical interpretation because there is no room for the concept of measurement 
in a fully quantized universe. These difficulties need a quantum mechanical remedy.
We might suppose a - yet hypothetic - universal decoherence mechanism
to make the wave function $\Psi(g,q)$  collapse into `pointer states' which are, like the semi-classical states, 
localized in $g$. The relativistic implementation of this project is completely missing. 
The non-relativistic task is tractable. We introduce the following Euclidean distance between
two mass densities $f(\rb)$ and $f'(\rb)$:
\begin{equation}\label{dist}
\Vert f-f'\Vert_G^2=G\int\int [f(\rb)-f'(\rb)][f(\sx)-f'(\sx)]\frac{\dd\rb\dd\sx}{\vert\rb-\sx\vert}~,
\end{equation}
where $G$ is the Newton constant. Based on this, a dynamical collapse theory was born long ago
\cite{Dio86,Dio87,Dio89,Pen96,Pen98,Pen04}. Each collapse violates Einstein's classical equivalence principle. 
I discuss how its restoration may lead to an emergent Newtonian interaction.

\section{Collapse from gravity}
At the heart of the collapse theory is the master equation \cite{Dio86,Dio87} to describe decoherence governed
by the distance (\ref{dist}):
\begin{equation}\label{mast}
\frac{\dd\ro}{\dd t}=-\frac{i}{\hbar}[\Ho,\ro]
                     -\la\frac{G}{2\hbar}\int\int[\fo(\rb),[\fo(\sx),\ro]]\frac{\dd\rb\dd\sx}{\vert\rb-\sx\vert}
               \equiv-\frac{i}{\hbar}[\Ho,\ro]+\la\Dc\ro~,
\end{equation}
where $\ro$ is the density matrix, $\Ho$ is the Hamiltonian of the system, and $\Dc$ is the decoherence term whose
coupling will be tuned by the number $\la$. 
By adding a suitable stochastic mechanism to the r.h.s., we get the collapse equation \cite{Dio89}: 
\begin{equation}\label{coll}
\frac{\dd\ro}{\dd t}=-\frac{i}{\hbar}[\Ho,\ro]+\la\Dc\ro-\frac{1}{\hbar}\int\{\fo-\foa,\ro\}\dphi\dd\rb~,
\end{equation}
where the auxiliary random field $\dphi$ has vanishing mean and a white-noise correlation:
\begin{equation}\label{dphidphi}
\mathbf{M}[\dphi(\rb,t)\dphi(\sx,u)]=\la\frac{\hbar G}{\vert\rb-\sx\vert}\delta(t-u)~.
\end{equation}
We do not include the Newtonian interaction 
\begin{equation}\label{HG}
\Ho_G=-\frac{G}{2}\int\int \frac{\fo(\rb)\fo(\sx)\dd\rb\dd\sx}{\vert\rb-\sx\vert}
\end{equation}
into $\Ho$ because we want to see if this term can - in some way - emerge from the collapse mechanism (\ref{coll}).

We may, for simplicity, consider the free motion of two (or more) rigid homogeneous balls of mass $M$ and radius $R$, 
with c.o.m. canonical variables are $\xbo_1,\pbo_1$ and $\xbo_2,\pbo_2$, resp. Then $\Ho=(\pbo_1+\pbo_2)/2M$ and
\begin{equation}\label{f}
\fo(\rb)=\frac{M}{V_R}\left[\theta(R-\vert \rb-\xbo_1\vert)+\theta(R-\vert \rb-\xbo_2\vert)\right]~,
\end{equation}
where $V_R=4\pi R^3/3$.
The collapse eq.~(\ref{coll}) yields pure states for long time, then the collapse and the kinetic terms balance each other, 
the wave function becomes a product state of localized pointer states:
\begin{equation}\label{psixy}
\psi(\xb_1,\xb_2)=\psi_1(\xb_1)\psi_2(\xb_2)~.
\end{equation}

\section{Pointer states, probe trajectories}
The dynamics of the pointer states $\psi_1,\psi_2$ is independent and only statistically correlated. Therefore we consider a
single probe first. In the limit where the spread $\sigma_\psi$ of $\xbo$ in the given pure state $\psik$ satisfies
$\sigma_\psi\ll R$, the collapse eqs.~(\ref{coll},\ref{dphidphi}) reduce \cite{Dio89}:
\begin{equation}\label{collball}
\frac{\dd\psik}{\dd t}=-\frac{i}{\hbar}\frac{\pbo^2}{2M}\psik
                 -\frac{\la}{2\hbar}M\omega_G^2(\xbo-\xba)^2\psik
                 +\frac{1}{\hbar}\Fb(\xbo-\xba)\psik~,
\end{equation}
where $\omega_G=GM/R^3$ and $\Fb$ is a random auxiliary force of vanishing mean and of white-noise isotropic correlation: 
\begin{equation}\label{FF}
\mathbf{M}[\Fb(t)\circ\Fb(u)]=\la\hbar M\omega_G^2\delta(t-u) \mathbf{1}~.
\end{equation}
The latter equation follows from eq.~(\ref{dphidphi}) if one compares the stochastic terms of both
eqs.~(\ref{coll}) and (\ref{collball}) which yields the relationship:
\begin{equation}\label{Fdphi}
\Fb=-\frac{M}{V_R}\int\nabla\dphi(\xba+\rb)\theta(R-r)\dd\rb~.
\end{equation}

Observe, however, that if $\Fb$ and $\dphi$ were real external force and field, resp.,  
they would then contribute to the Hermitian terms $(i/\hbar)\Fb\xbo\psik$ and $(-i/\hbar)\int\dphi\fo\dd\rb\psik$.
But they stand there without the $i$: the collapse needs the non-unitary mechanism. Yet, this is not the last word
about the `missing $i$'.  

The longtime asymptotic solution of the collapse eq.~(\ref{collball}) - when collapse and dynamics reach an equilibrium - 
is well known \cite{Dio88b}. The equilibrium wave packet becomes a complex Gaussian pointer state
\begin{equation}\label{psiinf}
\psi(\xb)=\frac{1}{\sqrt[4]{2\pi\sigma_\infty^2}}
          \exp\left(i\pba\xb-\sqrt{-2i}\left(\frac{\xb-\xba}{2\sigma_\infty}\right)^2\right)
\end{equation}
of stationary (squared) width $\sigma_\infty^2=2(\hbar/M\omega_G\sqrt{\la})$. The c.o.m. of the wave packet will perform
a sort of `Brownian' motion:
\begin{eqnarray}\label{dpdt}
\frac{\dd\pba}{\dd t}&=&\Fb~,\\
                 \label{dxdt}
\frac{\dd\xba}{\dd t}&=&\frac{\pba}{M}+\frac{2\sigma_\infty^2}{\hbar}\Fb~.
\end{eqnarray}
The trajectory of the c.o.m. is the classical inertial one apart from the `Brownian' motion which becomes ignorable 
for macroscopic masses. However, our `Brownian' motion is not the standard one because the relationship 
$\dd\xba/\dd t=\pba/M$ is violated by a diffusion term in (\ref{dxdt}) therefore the strict continuity of 
the spatial trajectory is lost. In classical physics this would be a fatal inconsistency, in quantum physics
we must cope with it. As for the eq.~(\ref{dpdt}), it yields standard Brownian momentum diffusion (friction is ignored). 

We make a crucial observation related to the `missing $i$'. The imaginary
force $i\Fb$ in the collapse eq.~(\ref{collball}) yields a real force $\Fb$ in eq.~(\ref{dpdt}).
Such miraculous transition $i\Fb\rightarrow\Fb$ does not hold in general but for the equilibrium
shape (\ref{psiinf}) of the wave function. This will nevertheless allow us to argue that  
the `Brownian' trajectories may - in a subtle way - induce the central attractive Newtonian force $g=-GM/r^2$ 
in the c.o.m. system of the object.

\section{Excursion: Ambient pressure from 'Brownian' trajectory}
As a very remote analogue, I would like to present the model of another non-standard `Brownian' motion where the presence and
the average value of a central compressing force can exactly be derived from the `Brownian' motion. 
Suppose our classical rigid ball of radius $R$ and mass $M$ is put into a thin gas of small rigid ball molecules that will
repeatedly hit the `Brownian' ball whose trajectory becomes a random broken line. You would not think that this broken
line encodes the value $P$ of the ambient pressure. Yet it does! 
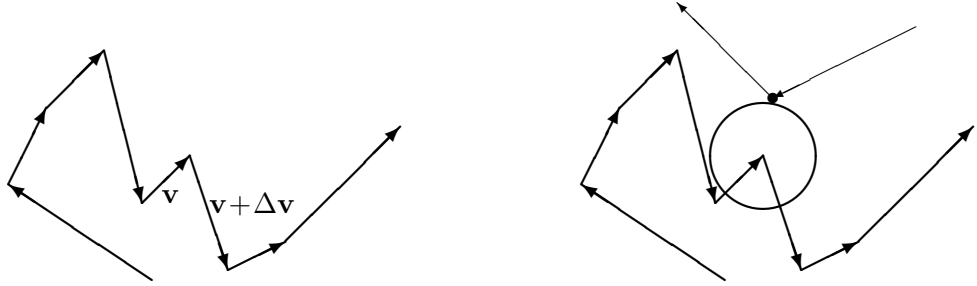
\begin{figure}[h]
\begin{center}
\setlength{\unitlength}{0.05in}
\begin{picture}(150,40)(0,0)
\thicklines
\put(30,10){\vector(-3,2){15}}
\put(15,20){\vector(1,2){4}}
\put(19,28){\vector(1,1){6}}
\put(25,34){\vector(1,-4){4}}
\put(29,18){\vector(1,1){5}}
\put(34,23){\vector(1,-3){4}}
\put(38,11){\vector(2,1){6}}
\put(44,14){\vector(1,1){12}}
\put(31,18){$\vb$}
\put(36,17){$\vb\!+\!\Delta\vb$}

\put(90,10){\vector(-3,2){15}}
\put(75,20){\vector(1,2){4}}
\put(79,28){\vector(1,1){6}}
\put(85,34){\vector(1,-4){4}}
\put(89,18){\vector(1, 1){5}}
\put(94,23){\vector(1,-3){4}}
\put(94,23){\circle{12}}
\put(95,29){\circle*{1}}
\thinlines
\put(95,29){\vector(-1, 1){10}}
\put(110,36.5){\vector(-2,-1){15}}
\thicklines
\put(98,11){\vector(2,1){6}}
\put(104,14){\vector(1,1){12}}
\end{picture}
\end{center}
\caption{The trajectory of a Brownian mass $M$ is visualized. The lines correspond to free 
inter-collision motion of the center of mass, break points are due to random separate 
collisions with the ambient molecules. On the r.h.s., one of the collisions is
detailed: an ambient molecule hits the surface of the Brownian mass to cause the 
momentum transfer $M\Delta\vb$.}
\end{figure}

Imagine that you are sitting inside the `Brownian' ball and experiencing the long sequence 
$\Delta\vb_1,\Delta\vb_2,\dots,\Delta\vb_k,\dots$ of velocity jumps during a long period $T$. 
This knowledge is not yet sufficient to conclude that there must be an ambient pressure. 
You have to know that you sit inside a rigid ball of mass 
$M$ and surface $4\pi R^2$ and the velocity jumps are caused by forces perpendicular to
the surface, which is indeed the case in all collisions, see Fig.~1. Then you argue that the absolute value 
$M\vert\Delta\vb_k\vert$ of each momentum change is caused by a force perpendicular to the surface.
The average of these perpendicular forces over a long period $T$ is thus equal to the average of
the experienced momentum jumps $M\vert\Delta\vb_k\vert$. On the other hand, the average of the perpendicular
forces is, by definition, equal to the ambient pressure times the surface, which leads you to the
exact relationship for the ambient pressure:
\begin{equation}\label{P}
P=\lim_{T\rightarrow\infty}\frac{M}{4\pi R^2}\sum_{k=1}^n \vert \Delta\vb_k\vert~.
\end{equation}

Can't we use the lessons of the above analogue to reach our original goal?
The first complete version \cite{Dio86} of the collapse theory \cite{Dio89} used jump-like
randomness instead of the diffusive term of the collapse eq.~(\ref{coll}), derived 
a collapse rule different from (\ref{collball}) in the special case of a rigid ball. 
Accordingly, the state satisfies the frictional Schr\"odinger equation:
\begin{equation}\label{dpsidrfr}
\frac{\dd\psik}{\dd t}=-\frac{i}{\hbar}\frac{\pbo^2}{2M}\psik
                 -\frac{\la}{2\hbar}M\omega_G^2(\xbo-\xba)^2\psik
                 +\frac{\la}{2\hbar}M\omega_G^2\sigma_\psi^2\psik
\end{equation}
apart from random jumps
\begin{equation}\label{jump}
\psik\longrightarrow\frac{\xbo-\xba}{\sigma_\psi}\psik
\end{equation}
which occur at the time dependent rate $\la M\omega_G^2\sigma_\psi^2$. These random jumps would yield a
broken line trajectory (with tiny discontinuities) similar to that on Fig.~1. Unfortunately, the equations of the jump-process
have not yet been solved. Therefore we return to the diffusive theory discussed in secs.~2 and 3. 

\section{Gravity from collapse?}
We wish to derive the presence of an average attractive Newtonian field $g=-GM/r^2$ from the features of
the random trajectory of the c.o.m. described by eqs.~(\ref{dpdt},\ref{dxdt}). 
Again, we could imagine that we are sitting inside the massive ball and experiencing the random accelerations 
which we shall attribute to the random force $\Fb$. This force is not yet the emergent Newtonian
interaction force we are looking for. Because of quantum features, the situation is more involved. 

Einstein's equivalence principle, that free falling objects move along geodesics, is violated by the collapse mechanism.
Can our diffusive trajectories (\ref{dpdt},\ref{dxdt}) become geodesics in a suitable curved space-time? 
No, they can not since the coordinate 
diffusion makes the spatial path discontinuous and this can never happen in any space-time continuum. 
Fortunately, the momentum diffusion of the trajectories can be attributed to curved space-time - in the
Newtonian limit - and this will lead us to the emergence of the Newtonian interaction.  

But we must learn some quantum subtleties first.
In fact, the c.o.m. trajectory $\{\xba,\pba\}$ of the probe is a mathematical fiction in a sense that 
it is not observable in a single run, only on a large statistics is $\{\xba,\pba\}$ testable. 
Fortunately enough, quantum mechanics tells us the \emph{unique observable}. Consider the following field:
\begin{equation}\label{phi}
\phi(\rb,t)=\dphi(\rb,t)-2\la G\int \frac{\langle\fo(\sx)\rangle_t}{\vert\rb-\sx\vert}\dd\sx
      \equiv\dphi(\rb,t)+2\la\phi^{scl}(\rb,t)~,
\end{equation}
where we have formally introduced the notation $\phi^{scl}$ for the $\psik$-dependent semi-classical Newtonian field (mean-field)
which does not yet play any dynamical role, which is still to play it after our arguments below.
Without going into the details of the continuous measurement interpretation \cite{Dio89,Dio88a,Dio90} of collapse,  
we invoke the fact that the above field $\phi$ is the only one available to the observer. 
Apart from $\psik$-independent functionals of the $\psik$-dependent $\phi$, 
no other quantities are testable. Testability (observability) has a perfect mathematical criterion. 
If we completed the r.h.s. of the collapse eq.~(\ref{coll}) by a deliberate $\psik$-dependent feed-back Hamiltonian, it would
not allow for the existence of the master equation like (\ref{mast}) while quantum mechanics requires its existence \cite{Gis84}.
The only guarantee of the existence is if we make the feed-back Hamiltonian depend on $\psik$ through the above $\phi$.

To fulfill or, at least, to approximate Einstein's classical equivalence principle,
we are looking for a curved space-time where the equilibrium c.o.m. behaviour of the probes (\ref{dpdt},\ref{dxdt}) will be as close to
the the classical geodesics as possible.  
In the Newtonian limit, our search for the space-time geometry means our search for a certain Newtonian potential
$\phi^?$ that we apply to the corresponding classical dynamics of the c.o.m.: 
\begin{eqnarray}\label{dpdtcl}
\frac{\dd\pba}{\dd t}&=&-\frac{M}{V_R}\int\nabla\phi^?(\xba+\rb)\theta(R-r)\dd\rb~,\\
                 \label{dxdtcl}
\frac{\dd\xba}{\dd t}&=&\frac{\pba}{M}~,
\end{eqnarray}
and we expect it yields similar dynamics to (\ref{dpdt},\ref{dxdt}) obtained from the collapse eq.~(\ref{coll}). 
We cannot expect the coincidence between the eqs.~(\ref{dxdt}) and (\ref{dxdtcl}) because coordinate diffusion is
a typical quantum effect. We expect the coincidence of the eqs.~(\ref{dpdt}) and (\ref{dpdtcl}). Let us compare
the forces on the r.h.s. of both, invoke eq.~(\ref{Fdphi}), and inspect that the choice $\phi^?=\dphi$ would match
the forces. Since, however, the field $\phi^?$ describes our space-time geometry  
therefore $\phi^?$ must be a testable field. The only choice is that $\phi^?$ is a $\psik$-independent functional of $\phi$.
The choice is uniqe:
\begin{equation}\label{choice}
\phi^?=\phi=\dphi+2\la\phi^{scl}~,
\end{equation}
otherwise the momentum diffusion in eq.~(\ref{dpdt}) remains unmatched.
For a single probe, the contribution of $\phi^{scl}$ cancels from the classical eq.~(\ref{dpdtcl}) so that it will,
as desired, coincide with the collapse result (\ref{dpdt}). However, in the presence of two (or more) probes,
$\phi^{scl}$ creates interaction between them: 
\begin{equation}\label{dpdtclscl}
\frac{\dd\pba}{\dd t}
=\Fb-2\la\frac{M}{V_R}\int\nabla\phi^{scl}(\xba+\rb)\theta(R-r)\dd\rb~.
\end{equation}
This corresponds to the classical Newtonian interaction, the standard strength will, however, require one half ($\la=1/2$) of the
decoherence strength  proposed by the author \cite{Dio86,Dio87,Dio89}; 
it corresponds to the choice by Penrose \cite{Pen96,Pen98,Pen04}.  

We are reaching our goal now. The result (\ref{dpdtclscl}) is still at variance with the collapse dynamics (\ref{dpdt}) because
of the Newtonian interaction term. One has to pay a price for the (partial) fulfillment of the equivalence principle, i.e., for the
gravitational explanation of the momentum diffusion caused by the collapse mechanism. We are happy to pay the price:
we add the Newtonian interaction $\Ho_G$ to the Hamiltonian part of the collapse eq.~(\ref{coll}) from the beginning.
As a result, the equilibrium (collapse vs. kinetics) c.o.m. trajectories of our quantum probes become classical geodesics - apart
from a tiny coordinate diffusion.    

\section{Discussion, outlook}
Ever since the birth of the gravity-related collapse models \cite{Dio86,Dio89}, the author can't stop thinking
about an emergent Newtonian interaction. The key concept has always been the re-enforcement of the Einstein
equivalence principle for the observable (testable) consequences of the quantum collapse. The present work
is a first tentative proposal, using heuristic as well as sophisticated theoretical arguments. The classical
corner stone of our construction is Einstein's equivalence principle known to everybody. The quantum corner stone is
known to few, that is the uniqueness of the $\psik$-dependent testable field in the theory of time-continuous
quantum measurement. The present work may leave concerns as to its ultimate consistence, it is nonetheless
a useful first attempt to outline the concept and some perspectives.

Some remarks are in order. First, it is obvious that without a topologically fluctuating space-time the coordinate
diffusion can never be interpreted classically. In this respect, our success is partial: the emergent space-time structure
restores momentum conservation rather than the full principle of equivalence. Second, the construction is sensitive
to the numeric strength factor $\la$ of decoherence. The value $\la=1$ allows for an underlying frictional Schr\"odinger-Newton equation 
\cite{Dio07,WezBri08}, the value $1/2$ does not. We desire this innocent looking discrepancy be relaxed by future refinements.   
Third, we imposed the classical equivalence principle on the c.o.m. trajectories $\{\xba,\pba\}$ which are not testable whereas
one corner stone of the construction is that the space-time should not depend but on testable quantity $\phi$. This is a 
conceptual flaw. An improved construction shall impose the classical equivalence principle to the testable quantities.
Fourth, the emergent Newtonian interaction assumes equilibrium pointer states for the probes. One might think that we
predict non-standard gravity for non-equilibrium wave functions which would be a basis for characteristic new phenomena to test.
This is not yet the case. If in a simple universe, dominated by rigid ball constituents, most of them are in
equilibrium pointer states than we predict standard Newton interaction everywhere.

\ack
This work was supported by the Hungarian OTKA Grant No. 49384 and No. 75129.


\vskip .3truecm
\begin{thebibliography}{99}
\bibitem{Whe68} Wheeler J A 1968 in {\it Battelle Rencontres: 1967 Lectures on Mathematical Physics} (New York: Benjamin)
\bibitem{DeW67} DeWitt B S 1967 {\it Phys. Rev.} {\bf 160} 1113
\bibitem{Per68} Peres A 1968 {\it Phys. Rev.} {\bf 171} 1335
\bibitem{Dio86} Di\'osi L 1986 {\it A quantum-stochastic gravitation model and the reduction of the wavefunction} 
        (in Hungarian) http://www.rmki.kfki.hu/~diosi/thesis1986.pdf
\bibitem{Dio87} Di\'osi L 1987 {\it Phys. Lett.} A {\bf 120} 377
\bibitem{Dio89} Di\'osi L 1989 {\it Phys. Rev.} A {\bf 40} 1165
\bibitem{Pen96} Penrose R 1996 {\it Gen. Relativ. Gravit.} {\bf 28} 581;
\bibitem{Pen98} Penrose R 1998 {\it Phil. Trans. R. Soc. Lond.} A {\bf 356} 1927
\bibitem{Pen04} Penrose R 2004 {\it The Road to Reality} (London: Jonathan Cape Publishers)
\bibitem{Dio88b} Di\'osi L 1988 {\it Phys. Lett.} A {\bf 132} 233
\bibitem{Dio88a} Di\'osi L 1988 {\it Phys. Lett.} A {\bf 129} 419
\bibitem{Dio90} Di\'osi L 1990 {\it Phys. Rev.} A {\bf 42} 5086
\bibitem{Gis84} Gisin N 1984 {\it Phys. Rev. Lett.} A {\bf 53} 1776
\bibitem{Dio07} Di\'osi L 2007 {\it J. Phys. Conf. Ser.}  {\bf 67} 011024
\bibitem{WezBri08} van Wezel J and van den Brink J 2008 {\it Philos. Mag.} {\bf88} 1659
\end{thebibliography}
\end{document}